\documentclass[twocolumn,showpacs,prl]{revtex4-1}
\usepackage{graphicx,amsmath,amssymb}
\usepackage{dcolumn,fancyhdr}
\usepackage{bm,natbib,mathrsfs}
\usepackage{times}
\usepackage{txfonts,palatino}
\usepackage{epstopdf}
\usepackage{latexsym}
\usepackage{epsfig}
\usepackage{bbold}
\usepackage[colorlinks=true,linkcolor=blue,urlcolor=blue,citecolor=blue]{hyperref}

\newcommand{\ket}[1]{\left\vert#1\right\rangle}

\newcommand{\bra}[1]{\left\langle#1\right\vert}

	\newcommand{\tr}[1]{\textrm{Tr} \left[ {#1} \right]} 

\begin{document}

\title{Work Extraction and Energy Storage in the Dicke Model}

\author{Lorenzo Fusco}
\affiliation{Centre for Theoretical Atomic, Molecular and Optical Physics, School of Mathematics and Physics, Queen's University, Belfast BT7 1NN, United Kingdom}
\author{Mauro Paternostro}
\affiliation{Centre for Theoretical Atomic, Molecular and Optical Physics, School of Mathematics and Physics, Queen's University, Belfast BT7 1NN, United Kingdom}
\author{Gabriele De Chiara}
\affiliation{Centre for Theoretical Atomic, Molecular and Optical Physics, School of Mathematics and Physics, Queen's University, Belfast BT7 1NN, United Kingdom}
\date{\today}

\begin{abstract}
We study work extraction from the Dicke model achieved using simple unitary cyclic transformations
keeping into account both a non optimal unitary protocol, and the energetic cost of creating the initial state. By analyzing the role of entanglement, we find that highly entangled states can be inefficient for energy storage when considering the energetic cost of creating the state. Such surprising result holds notwithstanding the fact that the criticality of the model at hand can sensibly improve the extraction of work. While showing the advantages of using a many-body system for work extraction, our results demonstrate that entanglement is not necessarily advantageous for energy storage purposes, when non optimal processes are considered. Our work shows the importance of better understanding the complex interconnections between non-equilibrium thermodynamics of quantum systems and correlations among their subparts.
\end{abstract}


\maketitle

In the last decades there has been a tremendous interest in the thermodynamical analysis of devices, in particular the conversion of heat into work and the extraction of work from a given substance, with a substantial effort in the study of quantum heat-engines, i.e. machines operating on a working medium given by a quantum substance \cite{nori,mahler,Friedenberger,Uzdin,campisipekola,campisisolo,kosloff,Kurizki,quan,AllahverdyanMahler,Bruno,Bruno2,Bruno3,Felix}. One of the applications of such a study is the possibility to identify strategies for the efficient storage of energy. One of the first steps towards the experimental realization of a quantum heat engine was made recently, with the theoretical proposal and demonstration of a single-ion heat engine in the classical regime \cite{Abah,Abah2,single}. Experiments studying the non-equilibrium thermodynamics of systems in the quantum regime have been realized recently with the scope of verifying the Jarzynski relation \cite{brazil,An}, and measuring entropy production resulting from processes implemented in quantum systems \cite{Papero}. \\
Whether or not quantum fluctuations and quantum correlations are effectively resources, when it comes to the efficiency of a heat-engine or to the maximization of extraction of work from a system, is still an open point. An enhancement of work extraction when using two-/three-qubit entangled working media has been shown experimentally~\cite{mauroextraction}. However the enhanced efficiency of work extraction from entangled states is effective only for small quantum systems \cite{Marti}.\\ For the case of a machine using a classical working medium on the verge of a phase transition, a boost in the efficiency has been predicted \cite{CampisiFazio}. On the other hand the study of many-body quantum heat-engines is still at its infancy \cite{DelCampo,fialko,Azimi,Poletti}, and we need to  understand whether a many-body quantum system can give an improvement in this respect as compared to a sequence of many heat-engines each operating with a single particle. Thus it is timely to proceed towards a systematic study of the properties of such devices. In particular the level of control over cold-atomic systems suggests that they could be extremely valuable as a test-bed for such devices. An emblematic example is given by the experimental realization of the Dicke model in an intracavity atomic system \cite{DickeETH}. The technology available at hand is mature enough to assess the thermodynamics of such cold-atomic system in the fully quantum regime. Also, the presence of a superradiant phase transition has been shown to play a role in the work output of a such an engine \cite{Hardal}.\\
\indent In this paper we take a significantly different approach with respect to previous studies. We quantify the relation between the energy extracted and the energy initially stored, for \emph{practical} cyclic processes: we characterize work extraction and energy storage, putting constrains on the optimality of the protocol motivated by the experimental control available over the system.
We then compare such practical protocols with the optimal ones.
We find that the quantum phase transition can improve the extraction of work. However, by considering the energetic cost of creating the initial state, and through an analysis of the role of entanglement, we show that highly entangled states can be inefficient for energy storage.  In particular our results show the existence of a non trivial link between non-equilibrium thermodynamics of quantum systems and entanglement for non-optimal unitarily operating devices.\\

\noindent \emph{Work Extraction Formalism -}
In the following we assume to drive cyclically the state of a quantum system with a time-dependent periodic Hamiltonian , with $\hat H(t)$ the instantaneous Hamiltonian of the system and  $t_i$ and $t_f$ the initial and final time of the evolution respectively, without contact to external reservoirs. Since work extraction from equilibrium state is forbidden by Thomson's formulation of the second law \cite{AllahverdyanThom}, we consider initial out-of-equilibrium states.\\
Let us suppose the initial state and the initial Hamiltonian to be given by $\hat{\rho}(t_i)=\sum_j r_j |r_j\rangle \langle r_j|$ and $\hat{H}(t_i)=\sum_j \epsilon_j \ket{\epsilon_j}\bra{\epsilon_j}$, where the ordering \mbox{$r_1\ge r_2\ge ...$}, and \mbox{$\epsilon_1\le \epsilon_2\le ...\,$} is assumed. Due to unitarity all of the eigenvalues of the initial state are preserved at any time.
The least energetic final state is $\hat{\rho}(t_f)_{\text{pass}}=\sum_j r_j |\epsilon_j\rangle \langle\epsilon_j|$. This final state commutes with the Hamiltonian $\hat{H}(t_i)$ and so is stationary, and the ordering of the eigenvalues is such that no work can be further extracted from it, making it passive. Associated to this optimal protocol we have the maximum extraction of work by the amount ${\cal E}=\sum_{ij}r_j \epsilon_i (|\langle r_j\ket{\epsilon_i}|^2-\delta_{ij})$, called ergotropy \cite{Allahverdyan2004}.

\noindent
\emph{Dicke Model -}
We consider the Dicke model: an emblematic model in quantum optics \cite{GrossHaroche}, also widely used as a benchmark for studying the behavior of quantum many-body systems with a quantum phase transition \cite{Vidal,Carmichael,PRLethDicke,Brennecke}.
 The Dicke Hamiltonian describes the coupling between an ensemble of $N$ two-level atoms and a single cavity mode and reads
\begin{equation}
\hat{H}=\omega_0 \hat{J}_z+\omega \hat{a}^{\dagger}\hat{a}+\frac{\lambda}{\sqrt{2j}}(\hat{a}+\hat{a}^{\dagger})(\hat{J}_+ + \hat{J}_-),
\end{equation}
where $\omega_0$ is the single atom two-level energy splitting, $\omega$ is the cavity frequency, and $\lambda$ is the atom-cavity interaction strength \cite{Dicke}. However the Dicke model is implemented experimentally with a hybrid cold-atomic system in an optical cavity \cite{DickeETH}, in which case the parameters must be interpreted differently as explained later in this article. The operators \mbox{$\hat{J}_i$ $(i=x,y,z,+,-)$} are collective angular momentum operators, with standard commutation relations, that allow to describe the atomic ensemble as a pseudo-spin of length \mbox{$j=N/2$}.
We can define the mean fields as \mbox{$\langle \hat{a}\rangle=\alpha$, $\langle \hat{J}_- \rangle=\beta$, $\langle \hat{J}_z \rangle=w$}, and write semiclassical equations of motion for them
derived from the Heisenberg equations, replacing operators with expectation values. 
The critical coupling \mbox{$\lambda_{\text{cr}}=\sqrt{\omega \omega_0}/2$} defines the separation point between the two fixed-point solutions of the semiclassical equations:
for \mbox{$\lambda<\lambda_{\text{cr}}$}, the so called normal phase, the mean fields are null; while for \mbox{$\lambda>\lambda_{\text{cr}}$}, the so called superradiant phase, both the atoms and field acquire macroscopic mean-fields of both signs.
With a standard Holstein-Primakoff transformation \mbox{$\hat{J}_+=\hat{b}^{\dagger}\sqrt{2j-\hat{b}^{\dagger}\hat{b}}$}, \mbox{$\hat{J}_-=\sqrt{2j-\hat{b}^{\dagger}\hat{b}}\hat{b}$}, \mbox{$\hat{J}_z=\hat{b}^{\dagger}\hat{b}-j$} \cite{Primakoff}, we can introduce the fluctuations operators $\delta\hat{a}=\hat{a}-{\alpha}$, $\delta \hat{b}=\hat{b}-\widetilde{{\beta}}/\sqrt{N}$,
where \mbox{$\alpha$} and \mbox{$\widetilde{\beta}= \langle \hat{b}\rangle$}
are chosen as the steady-state mean fields.
Explicitly, we get $(\hbar=1)$
\begin{equation}
\label{DickeDisplaced}
\begin{aligned}
\hat{H}&=\frac{\widetilde{\omega_0}}{2} \left(\hat{A}_x^2+\hat{A}_y^2\right)+\frac{\omega}{2}\left(\hat{P}_x^2+\hat{P}_y^2\right)+2\widetilde{\lambda}\hat{P}_x \hat{A}_x-2\mu \hat{A}_x^2 \\
&=\epsilon^{-}\hat{d}^{\dagger}\hat{d}+\epsilon^{+}\hat{c}^{\dagger}\hat{c}+\frac{1}{2}\Bigl(\epsilon^{-}+\epsilon^{+}-\omega-\widetilde{\omega_0}\Bigr),
\end{aligned}
\end{equation}
where the eigenvalues $\epsilon^+$ and $\epsilon^-$ and the coefficients $\widetilde{\omega_0}, \widetilde{\lambda}, \mu$ in Eq.~\eqref{DickeDisplaced} are reported in supplementary information (SI), and the quadrature operators are defined by 
$\hat{P}_x=\left(\delta\hat{a}^{\dagger}+\delta\hat{a}\right)/\sqrt{2}$, $\hat{P}_y=i\left(\delta\hat{a}^{\dagger}-\delta\hat{a}\right)/\sqrt{2}$,
$\hat{A}_x=\left(\delta\hat{b}^{\dagger}+\delta\hat{b}\right)/\sqrt{2}$, $\hat{A}_y=i\left(\delta\hat{b}^{\dagger}-\delta\hat{b}\right)/\sqrt{2}$.
In the last line of Eq.~\eqref{DickeDisplaced} we have introduced the polariton operators $\hat{d}$ and $\hat{c}$, that are connected to the local modes operators $\delta\hat{a}$ and $\delta\hat{b}$ via the matrix equation \mbox{$\bm{\delta \hat{a}}=\bm{M}\cdot\bm{\hat{d}}$}, where we have used the vector notation \mbox{$\bm{\delta\hat{a}}=(\delta\hat{a},\delta\hat{a}^{\dagger},\delta\hat{b},\delta\hat{b}^{\dagger})^T$}, and \mbox{$\bm{\hat{d}}=(\hat{d},\hat{d}^{\dagger},\hat{c},\hat{c}^{\dagger})^T$}.
Unless otherwise stated, in what follows we assume a constant value of the atomic frequency $\omega_0$.
For the experimental setup in Ref.~\cite{DickeETH}, the Dicke model is found as an effective Hamiltonian model describing a system that consists of a Bose-Einstein condensate (BEC) loaded into a high-finesse optical cavity, transversally pumped with a standing wave laser far-off resonant with respect to the atomic transition. In this case the mapping to the Dicke model is realised with an effective frequency $\omega$ given by the detuning between the pump frequency $\omega_p$ and the cavity mode frequency \mbox{$\omega_c$ ($\omega=\omega_p-\omega_c$)}. Thus $\omega$ is changed by varying the pump frequency.\\
The coupling parameter is given by \mbox{$\lambda=N/2 \sqrt{g_0 \Omega_p/\Delta_a}$}, where $g_0$ is the atom-cavity coupling, $\Omega_p$ is the pump Rabi frequency, and \mbox{$\Delta_a=\omega_p-\omega_a$} is the pump-atom detuning. A variation of $\lambda$ can be obtained with a quench of the intensity of the pump, controlled via $\Omega_p$. A variation of the pump frequency $\omega_p$ instead determines a variation of two parameters of the effective Dicke model, $\omega$ and $\lambda$. However, in order to realise an independent variation of $\omega$ such that it does not affect $\lambda$, we can realise simultaneously  two protocols \mbox{$\omega_{p_1}\rightarrow\omega_{p_2}$} and \mbox{$\Omega_{p_1}\rightarrow\Omega_{p_1}(\omega_{p_2}-\omega_a)/(\omega_{p_1}-\omega_a)$}. We thus assume the independent manoeuvrability of such parameters.
\noindent
\emph{Mean Field Contribution to the Work -}
The fixed points \mbox{$(\alpha_s,\beta_s)$} of the semiclassical equations are local minima of the mean energy $E:=\langle\hat{H}\rangle$, as a function of $\alpha$ and $\beta$. This means that the system starting slightly off the fixed point is in a classical non-equilibrium state and, according to Thomson's formulation of the second law, a cyclic variation of the parameters can determine a classical contribution to the work extraction.  However, as we are interested in studying the contribution to the work extraction coming from the quantum fluctuations of the system, we start the protocol from the fixed point, so that the extracted energy is only due to the quantum fluctuations.

\begin{figure}[b!]
\includegraphics[width=8.5cm]{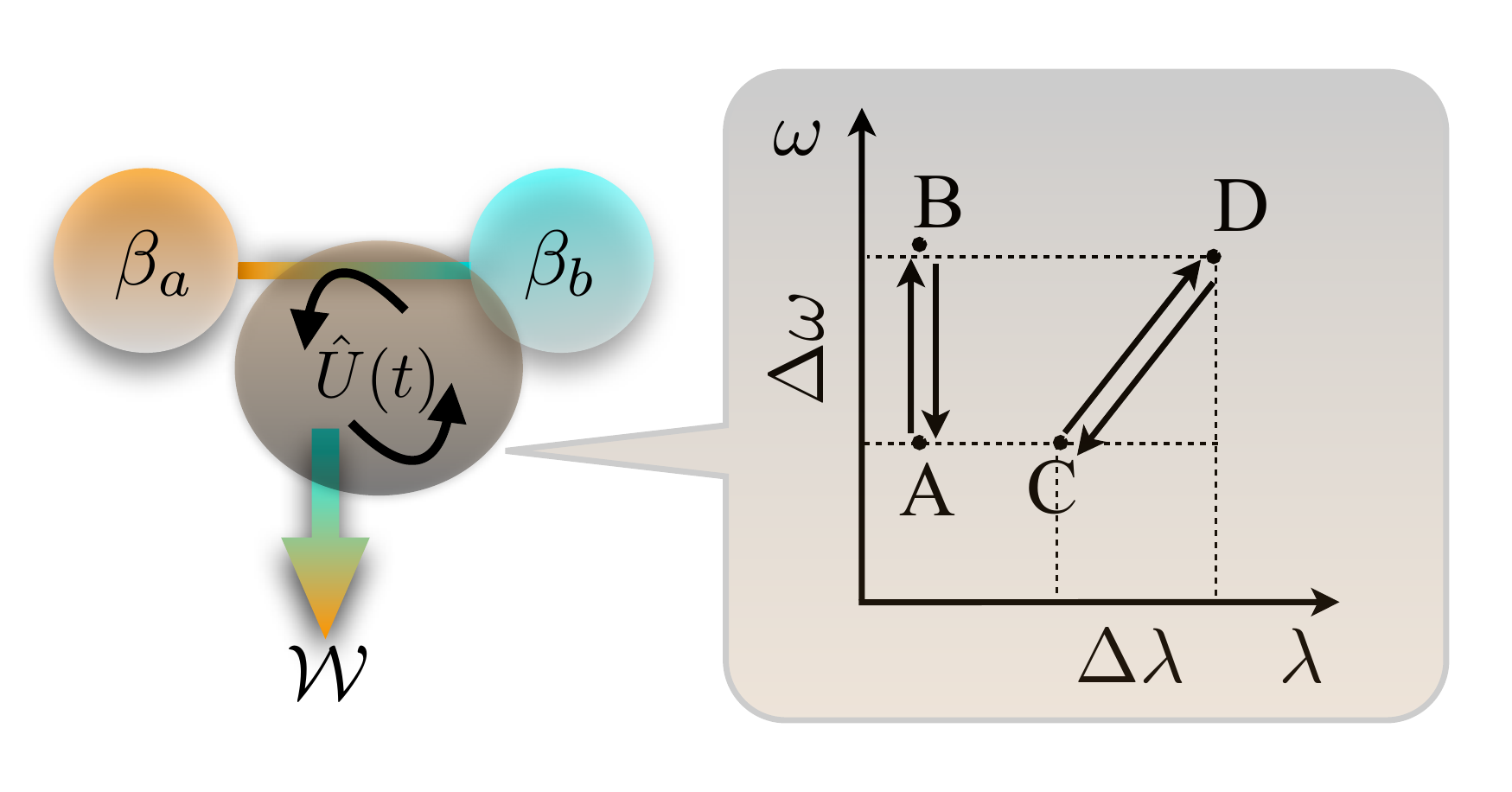}
\caption{Diagrammatic representation of the cycles. Initially the system is prepared in a locally thermal state, at inverse temperatures $\beta_a$ and $\beta_b$. The cyclic unitary transformation $\hat{U}(t)$ in the parameter space $(\lambda,\omega)$ is highlighted on the right: quench $A (C)\rightarrow B (D)$, evolution in  $B (D)$, quench $B (D)\rightarrow A (C)$, with a final extraction of work.} 
\label{Diagram} 
\end{figure}

\begin{figure*}[t!]
\includegraphics[width=17cm]{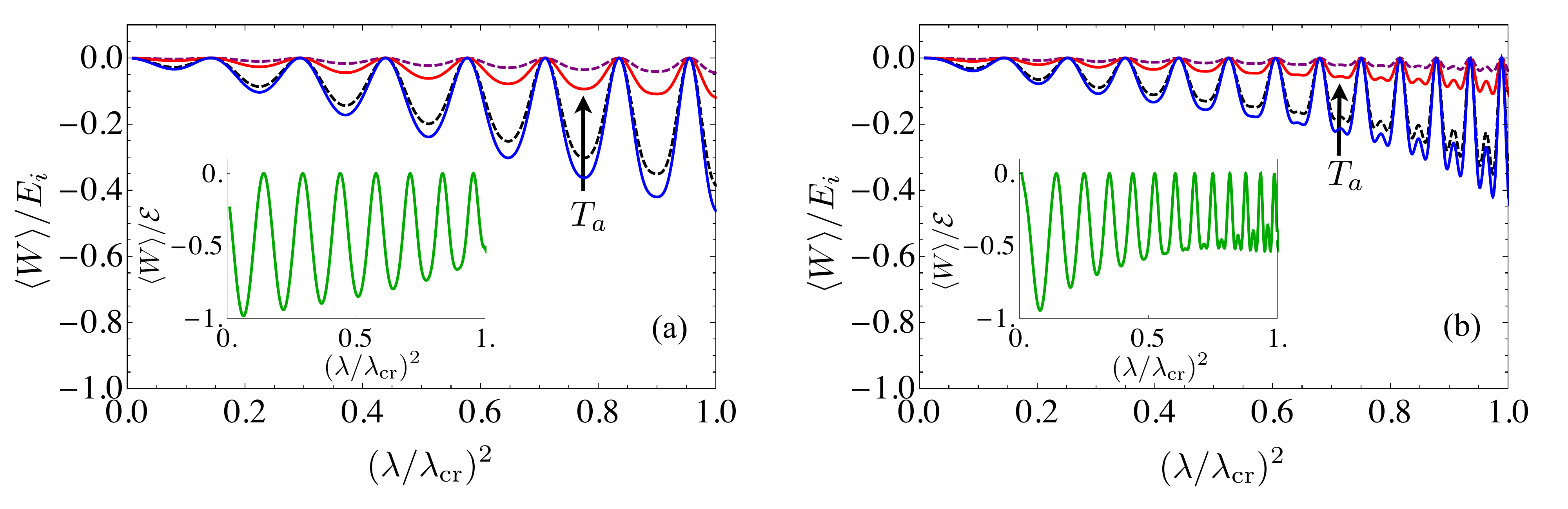}
\caption{Panel (a): Work-Energy ratio for the A-B cycle in Fig.~\ref{Diagram}, with \mbox{$\Delta\omega/2\pi=\omega/2\pi=15 \,\text{MHz}$} and \mbox{$\omega_0/2\pi=8.3\,\text{kHz}$}. Panel (b): Work-Energy ratio for the C-D cycle in Fig.~\ref{Diagram}, with $\Delta\omega=0.1 \,\omega$, $\Delta\lambda=-0.1 \,\lambda_{\text{cr}}$ . The free evolution time at point $B$ is \mbox{$\tau_B=0.003\,s$}. Dashed Purple: $\hat{\rho}^{\beta_a \beta_b}_1$, Solid Red: $\hat{\rho}^{\beta_a \beta_b}_2$, Dashed Black: $\hat{\rho}^{\beta_a \beta_b}_3$, Solid Blue: $\hat{\rho}^{\beta_a \beta_b}_4$, where $\beta_J=1/k_B T_J $, with $T_b=0.01K$, $T_a^1=10^{-1}K$, $T_a^2=10^{-1.5}K$, $T_a^3=10^{-2.5}K$, $T_a^4=10^{-3}K$. The green curve is the work-ergotropy ratio.}
\label{LocallyThermal}
\end{figure*}

Let us consider a time-dependent protocol, starting from a stationary value of the mean fields corresponding to a fixed point in the normal phase, i.e. for \mbox{$\lambda<\lambda_{\text{cr}}$}. In this case
the dynamics is such that, if we realise a general protocol so as to remain inside the normal phase, the mean fields will stay fixed at any instant of time.
If instead we realise a protocol that brings the system from the normal to the superradiant phase and then back, the mean fields will still remain fixed, but inside the superradiant phase this point is unstable.
Thus, for a very small change of the initial values there is an evolution of the mean fields, so that a work exchanged in the process must be positive for the reasons above mentioned. A positive contribution to work extraction coming from the mean fields would also be present for a cycle starting and ending inside the superradiant phase.

Therefore, for work extraction purposes that originate from quantum fluctuations, we have to limit ourselves to protocols within the normal phase. This result agrees with what was done in Ref.~\cite{silva}, where the authors limited the analysis to quenches inside the normal phase, since the case of crossing the two phases is not interesting from the point of view of the statistics of the work done, due to the macroscopic generation of excitations.
The cycles we consider can be schematised as follows:
i) preparation of the system in an initial state, ii) instantaneous quench \mbox{$\hat{H}_i\rightarrow \hat{H}_f$} (\mbox{$i=A$}, and \mbox{$f=B$} or \mbox{$i=C$}, and \mbox{$f=D$} with reference to Fig.~\ref{Diagram}) and evolution under \mbox{$\hat{H}_f$} for $t_f$, iii) instantaneous quench \mbox{$\hat{H}_f\rightarrow \hat{H}_i$}.
The average value of the work done in quenching the Hamiltonian can be written as \mbox{$\langle W \rangle=\bra{\psi(t_0)}(\hat{H}_H(t_f)-\hat{H}(t_i)) \ket{\psi(t_i)}$} \cite{LutzWork}, where $\hat{H}_H(t)=\hat{U}^{\dagger}(t)\hat{H}(t)\hat{U}(t)$ is the Hamiltonian in the Heisenberg representation, and $\hat{U}(t)$ is the evolution operator describing the process. We show in the supplementary information an analytical expression of the average work (see SI).\\
\noindent
\emph{Results for Locally Thermal States -}
In order to emulate, in our unitary framework, the effects of two thermal reservoirs, we consider the scenario sketched in Fig.~\ref{Diagram}.
An initial locally thermal state is prepared, where the two local modes are characterised by different inverse temperatures \mbox{$\beta_a=1/k_B T_a$} and \mbox{$\beta_b=1/k_B T_b$}, where $k_B$ is Boltzmann's constant, and the two oscillators are coupled; effectively realising Hamiltonian~\eqref{DickeDisplaced}.
After that, a cyclic unitary process is applied externally to the system, which can result eventually in an extraction of work.
We want to check whether the natural flow of energy, due to the particular initial state chosen here, can help us improve the extraction of energy.
We define locally thermal states
$\hat{\rho}^{\,\beta_a\,\beta_b}=\hat{\rho}^{\,\beta_a}\otimes \hat{\rho}^{\,\beta_b}$,
where the thermal states are \mbox{$\hat{\rho}^{\,\beta_j}=e^{-\beta_j \,\hat{H}_j}/{\cal Z}_j$}, with partition functions \mbox{${\cal Z}_j=\tr{\exp(-\beta_j\hat{H}_j)}$} $(j=a,b)$. The local Hamiltonians are $\hat{H}_a=\omega \, \delta\hat{a}^{\dagger}\delta\hat{a}$, $\hat{H}_b=\omega_0 \,\delta\hat{b}^{\dagger}\delta\hat{b}$.
The ergotropy for a locally thermal state in the polariton partition is zero
since, despite not being a thermal state,
it is however a passive state. The ergotropy for $\hat{\rho}^{\,\beta_a\,\beta_b}$
is instead not null, as shown in the supplementary information.
This results in the impossibility to extract energy from locally thermal states of non interacting systems, while this is possible for interacting systems.
This has motivated the study of the energetics of correlations in interacting systems \cite{Marti2}. \\
\indent In Fig.~\ref{LocallyThermal} we report the ratio between the total work and the average initial energy, for locally thermal states at different temperatures of the local mode $a$, as a function of the coupling parameter $\lambda$ which has been renormalized with respect to critical value of the initial Hamiltonian. This renormalisation causes a shift of the effective transition point for this figure of merit, since the Hamiltonian after the quench is characterised by a different value of the critical coupling. The work-energy ratio can be thought of as an efficiency of energy storage. 
In Fig.~\ref{LocallyThermal}~(a) and (b) we report the case of two-strokes cycles with $\Delta\omega= \omega$ (A-B cycle in Fig.~\ref{Diagram}),
and two-strokes cycles with $\Delta\omega=0.1\, \omega$ and $\Delta\lambda=-0.1 \,\lambda_{\text{cr}}$ (C-D cycle in Fig.~\ref{Diagram})
respectively. The green curves in the insets are the ratio between work and ergotropy.

\begin{figure*}[t!]
\includegraphics[width=17cm]{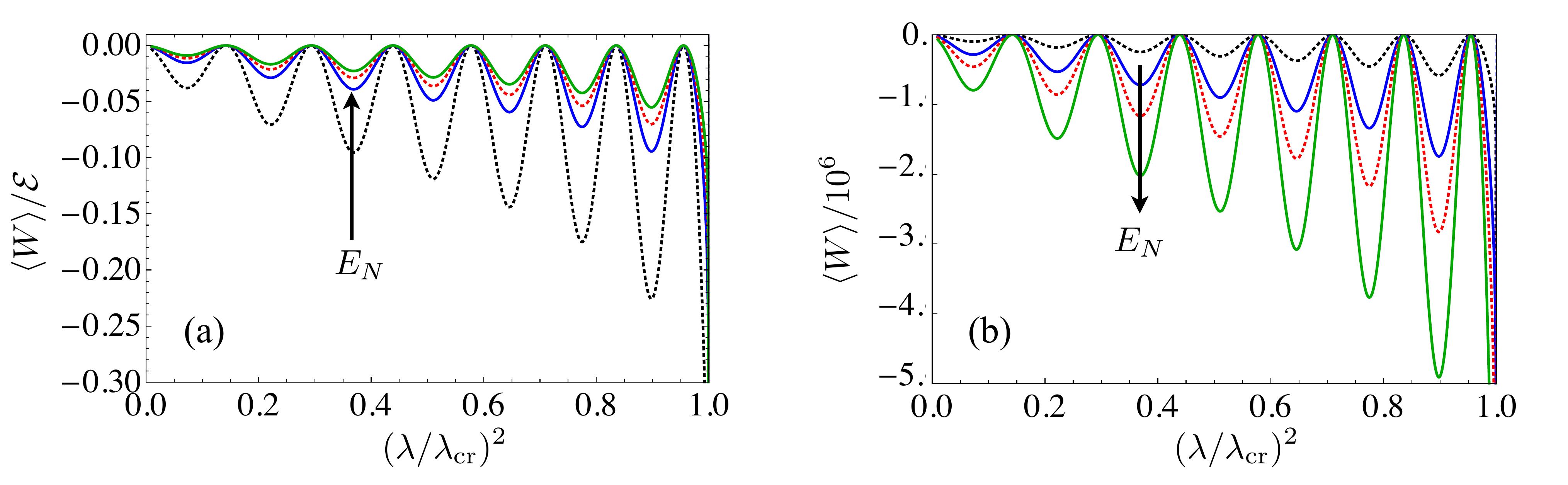}
\caption{Panel (a): Work-Ergotropy ratio for the locally passive entangled state in Eq.~\eqref{entdc} for the A-B cycle in Fig.~\ref{Diagram}, with \mbox{$\Delta\omega/2\pi= \omega/2\pi= \,15 \,\text{MHz}$}, and \mbox{$\omega_0/2\pi=8.3 \,\text{kHz}$}. Panel (b): total work for the same cycles as in panel~(a).}
\label{locallypassive}
\end{figure*}

We are in the extraction regime, witnessed by a negative sign of the work. The oscillations of the work, as a function of the coupling, are a consequence of the free evolution part of the cycle, and 
show the importance of choosing appropriate initial values of the coupling parameter to start the cycles from, in order to extract work from the system. For increasing temperature of one of the local modes, the fraction of work extracted to the initial energy decreases.
However the closer we are to the phase transition, the better the extraction of work is, as shown by the height of the negative peaks of the work-energy ratio. 
This shows how the presence of the phase transition helps retrieving the energy previously stored.\\ 
On the other hand the work-ergotropy ratio, reported in the insets of Fig.~\ref{LocallyThermal}, does not heavily depend on the temperature, for the particular regime of parameters considered, with values of the negative peaks $-1\le \langle W\rangle/{\cal E}\le -0.5$.
As the ergotropy is peaked around the phase transition, a non optimal process will be more inefficient close to the transition, and this is witnessed by the decreasing in absolute value by approaching the phase transition. This shows some of the consequences of the second law, inasmuch as despite being close to optimality (work-ergotropy close to one in absolute value) there is a fraction of the initial energy, spent to create the state, that we are not able to extract.



\noindent
\emph{Results for Locally Passive Entangled States -}
Previous studies have shown the importance of considering the role of quantum correlations for work extraction purposes, analysing for example the role of discord in work extraction from a $d$-level system \cite{Steve}. 
We now want to take into consideration the role of entanglement, and we are here interested in studying its role in the extraction of work for non optimal processes.
We evaluate the entanglement between the two modes via the logarithmic negativity of a two-mode Gaussian state \cite{PlenioNEG,Horodecki}.
We consider entangled states of the polaritonic modes that result in passive single-mode states. With this premises any work extraction can only be ascribed to entanglement.
Let us consider the state:
\begin{equation}
\label{entdc}
\ket{\psi_{dc}^{\text{ent}}}=\frac{1}{\sqrt{{\cal N}_{dc}}} \sum_{n=0}^{\infty}\exp{\left[-\beta(\epsilon_+ +\epsilon_-)n/4\right]}\ket{n}_d\ket{n}_c,
\end{equation}
with \mbox{${\cal N}_{dc}=\{1-\exp{\left[-\beta(\epsilon_+ +\epsilon_-)n/2\right]}\}^{-1}$}, whose marginals operators are
passive states.
The entanglement of state ~\eqref{entdc} does not heavily depend on the coupling $\lambda$ for our regime of parameters (see SI),
so that it allows us to use it as a free parameter.
In Fig.~\ref{locallypassive}~(a) we report the work-ergotropy ratio, and in Fig.~\ref{locallypassive}~(b) the total work, for two-strokes cycles.
If we were able to perform optimal work extraction, we would extract more work for more entangled states. This is shown to be true
also in the case of the non-optimal protocols considered here [cf. Fig.~\ref{locallypassive}~(b)]. However Fig.~\ref{locallypassive}~(a) shows that if we consider the fraction of extracted work to the maximum extractable, the behavior is reversed: the ratio is smaller for more entangled states. This behavior becomes more interesting if we consider that the figure of merit reported here is also an efficiency of energy storage, as for the initial pure state here chosen the ergotropy is equal to the average energy of the initial state. It must be noted that this is true even for the best case in which the state is a maximally entangled state (i.e. $\beta\to0$).

\noindent
\emph{Conclusions -}
We have shown a non trivial role played by  entanglement and quantum phase transitions, in the context of non equilibrium thermodynamics, for extraction and storing of energy, when considering both a non-optimal process and the energetic cost of creating the initial state. We have studied, for the emblematic example of the Dicke model, the advantages (and lack thereof) arising from the use of a many body quantum system as a working medium. Whenever we have the availability of a non optimal cyclic protocol, we have to choose favourable starting points for the cycle. If entanglement is the only resource for work extraction, the phase transition improves work extraction due to entanglement achieving a maximum at the phase transition. This intuition could lead us to prepare initially entangled states with high degrees of entanglement, to extract an increasing amount of work. However, we have shown that even for the best case in which the state approaches a maximally entangled state, the energy spent to create this state overcomes the gain in the possible extraction of work. Our results provide guidelines for the development of the new technology based on quantum machines.

We acknowledge insightful discussions with Cristiano Ciuti, Tobias Donner, Gianluca Francica, John Goold, Lorenz Hruby, Renate Landig, Rafael Mottl, and Simon Pigeon. This work was supported by the John Templeton Foundation (grant number 43467), the UK EPSRC (EP/J009776/1), the EU Collaborative Project TherMiQ (Grant Agreement 618074), and the Julian Schwinger Foundation (JSF-14-7-0000). Part of this work was supported by COST Action MP1209 "Thermodynamics in the quantum regime".


\newpage
\onecolumngrid
\renewcommand{\bibnumfmt}[1]{[S#1]}
\renewcommand{\citenumfont}[1]{S#1}
\section*{Supplementary Information}
\vspace{1cm}
\twocolumngrid

\renewcommand{\theequation}{S-\arabic{equation}}
\setcounter{equation}{0}  
\section{Diagonalization of the Dicke Hamiltonian}  
\label{prima}



Here we show the details of the diagonalization of the Dicke Hamiltonian. The parameters of the Hamiltonian are given by
\begin{equation}
\begin{aligned}
\label{parameterss}
\widetilde{\omega_0}&=\omega_0-\frac{2\lambda\alpha_s{\beta_s}}{N^{3/2}\sqrt{1-\frac{{\beta_s}^2}{N^2}}},\\
\mu&=\frac{\lambda {\alpha_s}{\beta_s}}{N^{3/2}\sqrt{1-\frac{{\beta_s}^2}{N^2}}}\left(1+\frac{{\beta_s}^2}{2(N^2-{\beta_s}^2)}\right),\\
\widetilde{\lambda}&=\lambda \frac{1-2\frac{{\beta_s}^2}{N^2}}{\sqrt{1-\frac{{\beta_s}^2}{N^2}}},\\
E_0&=\omega \alpha_s^2+\omega_0\left(\frac{{\beta_s}^2}{N}-\frac{N}{2}\right)+4\lambda \frac{{\alpha_s\beta_s}}{\sqrt{N}}\sqrt{1-\frac{{\beta_s}^2}{N^2}},
\end{aligned}
\end{equation}
where the steady-state mean fields are

\begin{equation}
\label{alphabeta3}
\alpha_s=
\begin{cases}
\begin{array}{c c}
0\hspace{0.4cm}&\text{for}\hspace{0.4cm}\lambda<\lambda_{\text{cr}},\\
\mp\frac{\lambda \sqrt{N}}{\omega}\sqrt{1-\left(\frac{\lambda_{\text{cr}}}{\lambda}\right)^4}\hspace{0.4cm}&\text{for}\hspace{0.4cm}\lambda>\lambda_{\text{cr}},\\
\end{array}
\end{cases}
\end{equation}
and
\begin{equation}
\label{alphabeta4}
\beta_s=
\begin{cases}
\begin{array}{c c}
0\hspace{0.4cm}&\text{for}\hspace{0.4cm}\lambda<\lambda_{\text{cr}},\\
\pm\frac{N}{2} \sqrt{1-\left(\frac{\lambda_{\text{cr}}}{\lambda}\right)^4}\hspace{0.4cm}&\text{for}\hspace{0.4cm}\lambda>\lambda_{\text{cr}}.\\
\end{array}
\end{cases}
\end{equation}

Then we apply a transformation that renormalizes the effective masses of the oscillators by going into the phase space \cite{emary}
\begin{equation}
\label{anncrea}
\begin{aligned}
&\hat{x}=\frac{1}{\sqrt{2\omega}}(\delta\hat{a}^{\dagger}+\delta\hat{a}), \hspace{0.5cm} \hat{p}_x=i\sqrt{\frac{\omega}{2}}(\delta\hat{a}^{\dagger}-\delta\hat{a}),\\
&\hat{y}=\frac{1}{\sqrt{2\widetilde{\omega_0}}}(\delta\hat{b}^{\dagger}+\delta\hat{b}), \hspace{0.5cm} \hat{p}_y=i\sqrt{\frac{\widetilde{\omega_0}}{2}}(\delta\hat{b}^{\dagger}-\delta\hat{b}).
\end{aligned}
\end{equation}
After this transformation we get
\begin{equation}
\begin{aligned}
\hat{H}&=\frac{1}{2}\Biggl\{ \omega^2 \hat{x}^2 +\hat{p}_x^2+ (\widetilde{\omega_0}^2-4\mu\widetilde{\omega_0}) \hat{y}^2 +\hat{p}_y^2+\\
&+4\widetilde{\lambda} \sqrt{\omega \widetilde{\omega_0}}\hat{x}\hat{y}-\widetilde{\omega_0}-\omega\Biggr\}+E_0.
\end{aligned}
\end{equation}
Then we rotate the system coordinate with the transformation (we will indicate the Bogoliubov angle as $\gamma^{(B)}$)
\begin{equation}
\hat{x}=\hat{q}_1 \cos \gamma^{(B)}+\hat{q}_2 \sin \gamma^{(B)}, \hspace{0.1cm} \hat{y}=-\hat{q}_1 \sin \gamma^{(B)}+\hat{q}_2 \cos \gamma^{(B)}
\end{equation}
and similar transformations apply to the momentum operators. In the new representation the interaction is removed if we choose the angle $\gamma^{(B)}$ such that
\begin{equation}
\tan (2 \gamma^{(B)})=\frac{4\widetilde{\lambda} \sqrt{\omega \widetilde{\omega_0}}}{\widetilde{\omega_0}^2-4\mu\widetilde{\omega_0}-\omega^2}.
\end{equation}
The Hamiltonian for the two decoupled oscillators is
\begin{equation}
\hat{H}=\frac{1}{2}\Biggl\{ \epsilon^{-} \hat{q}_1^2+\hat{p}_1^2+\epsilon^{+} \hat{q}_2^2+\hat{p}_2^2  -\omega-\widetilde{\omega_0}\Biggr\}+E_0,
\end{equation}
where the energies are
\begin{equation}
\label{eigenmodes}
\epsilon^{\pm}=\sqrt{\frac{1}{2}\Biggl(z+2\omega^2\pm\text{sign} \left(z\right) \sqrt{z^2+16\widetilde{\lambda}^2 \omega \widetilde{\omega_0}}\Biggr)},
\end{equation}
with $z=\widetilde{\omega_0}^2-4\mu\widetilde{\omega_0}-\omega^2$. Then again we apply the transformation
\begin{equation}
\begin{aligned}
&\hat{q}_1=\frac{1}{\sqrt{2\epsilon^{-}}}(\hat{d}^{\dagger}+\hat{d}), \hspace{0.5cm} \hat{p}_1=i\sqrt{\frac{\epsilon^{-}}{2}}(\hat{d}^{\dagger}-\hat{d}),\\
&\hat{q}_2=\frac{1}{\sqrt{2\epsilon^{+}}}(\hat{c}^{\dagger}+\hat{c}), \hspace{0.5cm} \hat{p}_2=i\sqrt{\frac{\epsilon^{+}}{2}}(\hat{c}^{\dagger}-\hat{c}),
\end{aligned}
\end{equation}
and we end up finally with the Hamiltonian
\begin{equation}
\hat{H}=\epsilon^{-}\hat{d}^{\dagger}\hat{d}+\epsilon^{+}\hat{c}^{\dagger}\hat{c}+\frac{1}{2}\Bigl(\epsilon^{-}+\epsilon^{+}-\omega-\widetilde{\omega_0}\Bigr)+E_0
\end{equation}

In the phase-space the diagonalization is obtained with the transformation \mbox{$\bm{\delta \hat{a}}=\bm{M}\cdot\bm{\hat{d}}$}, with \mbox{$\bm{\delta\hat{a}}=(\delta\hat{a},\delta\hat{a}^{\dagger},\delta\hat{b},\delta\hat{b}^{\dagger})^T$} and \mbox{$\bm{\hat{d}}=(\hat{d},\hat{d}^{\dagger},\hat{c},\hat{c}^{\dagger})^T$}. The symplectic matrix $\bm{M}$ is
\begin{equation}\label{M}
\bm{M} = \begin{pmatrix}
A_+&A_-&B_+&B_-\\
A_-&A_+&B_-&B_+\\
C_+&C_-&D_+&D_-\\
C_-&C_+&D_-&D_+\\
\end{pmatrix},
\end{equation}
where the coefficients are
\begin{equation}
\begin{aligned}
A_{\pm}&=\frac{1}{2}\cos \left(\gamma^{(B)}\right)\left(\sqrt{\frac{\omega}{\epsilon_-}}\pm\sqrt{\frac{\epsilon_-}{\omega}}\right),\\
B_{\pm}&=\frac{1}{2}\sin \left(\gamma^{(B)}\right)\left(\sqrt{\frac{\omega}{\epsilon_+}}\pm\sqrt{\frac{\epsilon_+}{\omega}}\right),\\
C_{\pm}&=-\frac{1}{2}\sin \left(\gamma^{(B)}\right)\left(\sqrt{\frac{\widetilde{\omega_0}}{\epsilon_-}}\pm\sqrt{\frac{\epsilon_-}{\widetilde{\omega_0}}}\right),\\
D_{\pm}&=\frac{1}{2}\cos \left(\gamma^{(B)}\right)\left(\sqrt{\frac{\widetilde{\omega_0}}{\epsilon_+}}\pm\sqrt{\frac{\epsilon_+}{\widetilde{\omega_0}}}\right).\\
\end{aligned}
\end{equation}

\section{Sudden Quench Cycle}  
\label{suddenstrokes}
In the case of a cycle in which the strokes are realised with successive sudden quenches, it is possible to obtain analytical expressions for the average work. In what follows we will consider explicitly the case of a pure initial state, just for convenience of calculation, but everything can be easily transposed to the case of a general mixed initial state with the proper averages taken.\\
 Suppose that we want to realise a four strokes cycle (A-B-C-D), and we initially prepare the state of the system in the state $\ket{\psi_A}$ 
where $A$ labels the starting point of the cycle in the parameters space. For a sudden quench the unitary evolution operator is the identity \mbox{$\hat{U}(t)=\hat{\mathbb{1}}$}, so that for the average work we have
\begin{equation}
\label{workAB}
\begin{aligned}
&\langle W \rangle_{AB}=\bra{\psi_A} \left( \hat{H}_B - \hat{H}_A \right) \ket{\psi_A}={E_0}_B-{E_0}_A\\
&+\langle\epsilon_B^{-}\hat{d}^{\dagger}_B\hat{d}_B+\epsilon_B^{+}\hat{c}^{\dagger}_B\hat{c}_B\rangle-\langle\epsilon_A^{-}\hat{d}^{\dagger}_A\hat{d}_A+\epsilon_A^{+}\hat{c}^{\dagger}_A\hat{c}_A\rangle\\
&+\frac{1}{2}(\epsilon_B^{-}-\epsilon_A^{-}+\epsilon_B^{+}-\epsilon_A^{+}-\omega_B+\omega_A-\widetilde{\omega_0}_B+\widetilde{\omega_0}_A).
\end{aligned}
\end{equation}
In order to calculate this expression we use the relation between mode operators at different points in the parameter space
\begin{equation}
\bm{\hat{d}}_B=\bm{M}_B^{-1}\bm{M}_A\bm{\hat{d}}_A+\bm{M}_B^{-1}(\bm{\alpha}_A-\bm{\alpha}_B),
\end{equation}
that allows us to express the terms \mbox{$\hat{d}_B^{\dagger}\hat{d}_B$} and \mbox{$\hat{c}_B^{\dagger}\hat{c}_B$} in terms of operators $\bm{\hat{d}}_A$. It is supposed that we know the covariance matrix
\begin{equation}
\label{covarianceMatrix}
(\bm{\sigma}_{A}^{d})_{ij} =\frac{1}{2}\langle (\bm{\hat{d}}_A)_i(\bm{\hat{d}}_A)_j+(\bm{\hat{d}}_A)_j(\bm{\hat{d}}_A)_i\rangle\,,
\end{equation}
where the indices $i$ and $j$ denote the components of the respective vectors or matrix. In what follows we use the convention that number as indices denote elements of vectors or matrices, while letters as indices denote different points in the parameters space. If we indicate with $\bm{\hat{d}}_{B\,i}$ the $i$-th element of vector $\bm{\hat{d}}_B$, and similarly for others, we have
\begin{equation}
\begin{aligned}
\hat{d}_B^{\dagger}\hat{d}_B&=\Bigl(\bm{M}_B^{-1}\bm{M}_A\bm{\hat{d}}_A+\bm{M}_B^{-1}(\bm{\alpha}_A-\bm{\alpha}_B)\Bigr)_2 \\
&\times\Bigl(\bm{M}_B^{-1}\bm{M}_A\bm{\hat{d}}_A+\bm{M}_B^{-1}(\bm{\alpha}_A-\bm{\alpha}_B)\Bigr)_1\\
\hat{c}_B^{\dagger}\hat{c}_B&=\Bigl(\bm{M}_B^{-1}\bm{M}_A\bm{\hat{d}}_A+\bm{M}_B^{-1}(\bm{\alpha}_A-\bm{\alpha}_B)\Bigr)_4 \\
&\times\Bigl(\bm{M}_B^{-1}\bm{M}_A\bm{\hat{d}}_A+\bm{M}_B^{-1}(\bm{\alpha}_A-\bm{\alpha}_B)\Bigr)_3.\\
\end{aligned}
\end{equation}
Given the covariance matrix $\bm{\sigma}^d_A$ of the initial state $|\psi_A\rangle$, we can conveniently express everything in terms of elements of the matrix \mbox{$\bm{Q}^{AB}=\bm{M}_B^{-1}\bm{M}_A$} and vector \mbox{${\bm V}^{AB}=\bm{M}_B^{-1}(\bm{\alpha}_A-\bm{\alpha}_B)$} as follows:

\begin{equation}
\bm{\hat{d}}_B=\bm{Q}^{AB}\bm{\hat{d}}_A+\bm{V}^{AB},
\end{equation}


\begin{equation}
\label{DTDB}
\langle\psi_A|\hat{d}_B^{\dagger}\hat{d}_B|\psi_A\rangle=\sum_{ij}\bm{Q}_{2i}^{AB}\bm{Q}_{1j}^{AB}[(\bm{\sigma}^d_A)_{ij}+\bm{\Lambda}_{ij}]+\bm{V}_2^{AB}\bm{V}_1^{AB}
\end{equation}
\begin{equation}
\label{CTCB}
\langle\psi_A|\hat{c}_B^{\dagger}\hat{c}_B|\psi_A\rangle=\sum_{ij}\bm{Q}_{4i}^{AB}\bm{Q}_{3j}^{AB}[(\bm{\sigma}^d_A)_{ij}+\bm{\Lambda}_{ij}]+\bm{V}_4^{AB}\bm{V}_3^{AB},
\end{equation}


so that the work is given by
\begin{equation}
\begin{aligned}
&\langle W \rangle_{AB}=\epsilon_B^{-}\langle\hat{d}^{\dagger}_B\hat{d}_B\rangle+\epsilon_B^{+}\langle\hat{c}^{\dagger}_B\hat{c}_B\rangle-\epsilon_A^{+}[(\bm{\sigma}^d_A)_{43}+\bm{\Lambda}_{43}]\\
&-\epsilon_A^{-}[(\bm{\sigma}^d_A)_{21}+\bm{\Lambda}_{21}]+\Delta C_{AB},
\end{aligned}
\end{equation}
where the first two terms are given in Eqs.~\eqref{DTDB} and \eqref{CTCB}, and $\Delta C_{AB}$ account for the total constant part in Eq.~\eqref{workAB}.

For a second stroke (\mbox{$B\rightarrow C$}) we need to evaluate the following expression
\begin{equation}
\label{WBC}
\begin{aligned}
&\langle W \rangle_{BC}=\bra{\psi_A} e^{i\hat{H}_B \tau_B} \left( \hat{H}_C - \hat{H}_B \right)e^{-i\hat{H}_B \tau_B}  \ket{\psi_A}=\\
&=\langle\psi_A|e^{i\hat{H}_B \tau_B} \Bigl(\epsilon_{C}^{-}\hat{d}^{\dagger}_C\hat{d}_C+\epsilon_{C}^{+}\hat{c}^{\dagger}_C\hat{c}_C\Bigr)e^{-i\hat{H}_B \tau_B}|\psi_A\rangle\\
&-\langle \psi_A|\epsilon_{B}^{-}\hat{d}^{\dagger}_B\hat{d}_B+\epsilon_{B}^{+}\hat{c}^{\dagger}_B\hat{c}_B|\psi_A\rangle+\Delta C_{BC}.
\end{aligned}
\end{equation}
It is convenient to define the diagonal matrix
\begin{equation}
\bm{D}^B=\text{diag}\left(e^{-i \epsilon_B^{-}\tau_B},e^{i \epsilon_B^{-}\tau_B},e^{-i \epsilon_B^{+}\tau_B},e^{i \epsilon_B^{+}\tau_B}\right),
\end{equation}
so that we can write the evolution of the vector \mbox{$\bm{\hat{d}}_K\,(K=A,B,C,D)$} in matrix notation as
\begin{equation}
e^{i\hat{H}_K \tau_K}\bm{\hat{d}}_K e^{-i\hat{H}_K \tau_K}=\bm{D}^K \bm{\hat{d}}_K.
\end{equation}
With this definition we can compute the first two terms of Eq.~\eqref{WBC} as
\begin{equation}
\label{DTDC}
\begin{aligned}
&\langle\psi_A|e^{i\hat{H}_B \tau_B}\hat{d}_C^{\dagger}\hat{d}_C e^{-i\hat{H}_B \tau_B}|\psi_A\rangle=\\
&=\bra{\psi_A}(\bm{R}^{AC}\bm{\hat{d}}_A+\bm{S}^{AC})_2(\bm{R}^{AC}\bm{\hat{d}}_A+\bm{S}^{AC})_1\ket{\psi_A}
\end{aligned}
\end{equation}
and
\begin{equation}
\label{CTCC}
\begin{aligned}
&\langle\psi_A|e^{i\hat{H}_B \tau_B}\hat{c}_C^{\dagger}\hat{c}_C e^{-i\hat{H}_B \tau_B}|\psi_A\rangle=\\
&=\bra{\psi_A}(\bm{R}^{AC}\bm{\hat{d}}_A+\bm{S}^{AC})_4(\bm{R}^{AC}\bm{\hat{d}}_A+\bm{S}^{AC})_3\ket{\psi_A},
\end{aligned}
\end{equation}
with matrix \mbox{$\bm{R}^{AC}=\bm{Q}^{BC}\bm{D}^B\bm{Q}^{AB}$}, and vector \mbox{$\bm{S}^{AC}=\bm{Q}^{BC}\bm{D}^B\bm{V}^{AB}+\bm{V}^{BC}$}.
The meaning of expression for the matrix $\bm{R}^{AC}$ is straightforward. The matrix $\bm{Q}^{AB}$ is responsible for the connection between operators of points $A$ and $B$ in the parameter space due to the quench \mbox{$A\rightarrow B$}. Then, matrix $\bm{D}^B$ expresses the time evolution of the system at point $B$, and finally again matrix $\bm{Q}^{BC}$ realises the quench \mbox{$B\rightarrow C$}. Vector $\bm{S}^{AC}$ instead expresses the contribution coming from the mean fields in the evolution from $A$ to $C$. There can be a contribution from the difference of mean fields between $A$ and $B$ ($\bm{V}^{AB}$), then an evolution in B ($\bm{D}^B$) and finally a quench \mbox{$B\rightarrow C$} ($\bm{Q}^{BC}$); in addition there is also a contribution coming from the difference between the mean fields of $B$ and $C$.

The crucial point is that Eqs.~\eqref{DTDC} and \eqref{CTCC} are totally equivalent to Eqs.~\eqref{DTDB} and \eqref{CTCB}, so that we can use the same results in the latter expressions to evaluate the former ones, with the substitutions \mbox{$\bm{Q}^{AB}\rightarrow \bm{R}^{AC}$} and \mbox{$\bm{V}^{AB}\rightarrow \bm{S}^{AC}$}. The second term in Eq.~\eqref{WBC} for the work $\langle W\rangle_{BC}$ has already been evaluated for the work $\langle W\rangle_{AB}$.
If we keep on calculating the averages of work for each stroke in the same way, eventually we need to sum all the contributions to get the total average work for the cycle , which e.g. in the case of a 4-strokes cycle gives
\mbox{$\langle W\rangle_{\text{tot}}=\langle W\rangle_{AB}+\langle W\rangle_{BC}+\langle W\rangle_{CD}+\langle W\rangle_{DA}.\\$}

\section{Ergotropy for locally thermal states}

The ergotropy for a locally thermal state $\hat{\rho}^{\,\beta_d\, \beta_c}$ in the polariton partition is:
\begin{equation}
\label{ergo0}
{\cal E}\left(\hat{\rho}^{\,\beta_d\, \beta_c}\right)=E\left(\hat{\rho}^{\,\beta_d\, \beta_c}\right)-E\left(\hat{\rho}_{\text{pass}}^{\,\beta_d\, \beta_c}\right)=0,
\end{equation}
where \mbox{$E(\hat{\rho})$} denotes the average energy of the state $\hat{\rho}$ since, despite not being a thermal state because of the different local temperatures of the polariton modes, it is however a passive state. The ergotropy of the locally thermal state $\hat{\rho}^{\,\beta_a\, \beta_b}$ defined in the main text is instead:
\begin{equation}
\begin{aligned}
\label{ergoTaTb}
&{\cal E}\left(\hat{\rho}^{\,\beta_a\, \beta_b}\right)=E\left(\hat{\rho}^{\,\beta_a\, \beta_b}\right)-E\left(\hat{\rho}_{\text{pass}}^{\,\beta_a\, \beta_b}\right)=\\
&=\epsilon_c\left(\langle \hat{c}^{\dagger}\hat{c}\rangle_{\beta_a \,\beta_b}-\langle n^T_b\rangle\right)+\epsilon_d\left(\langle\hat{d}^{\dagger}\hat{d}\rangle_{\beta_a \,\beta_b}-\langle n^T_a\rangle\right)\\
\end{aligned}
\end{equation}
where
\begin{equation}
\langle n^T_a\rangle=\frac{1}{e^{\beta_a \omega}-1},\,\hspace{1cm} \langle n^T_b\rangle=\frac{1}{e^{\beta_b \omega_0}-1},
\end{equation}
and 
\begin{equation}
\begin{aligned}
\langle \hat{c}^{\dagger}\hat{c}\rangle_{\beta_a \,\beta_b}&= \tr{\hat{c}^{\dagger}\hat{c}\,\hat{\rho}^{\,\beta_a\, \beta_b}} \ne  \langle n^T_b\rangle\\
\langle \hat{d}^{\dagger}\hat{d}\rangle_{\beta_a \,\beta_b}&=\tr{\hat{d}^{\dagger}\hat{d}\,\hat{\rho}^{\,\beta_a\, \beta_b}} \ne \langle n^T_a\rangle,\\
\end{aligned}
\end{equation}
making the ergotropy of state $\hat{\rho}^{\,\beta_a\, \beta_b}$ different from zero. In fact, given the expression for the covariance matrix $\bm{\sigma}^{\,\beta_a\,\beta_b}_{ab}$ of state $\hat{\rho}^{\,\beta_a\,\beta_b}$ in the phase space basis $\bm{\delta\hat{a}}$
\begin{equation}
\bm{\sigma}^{\,\beta_a\,\beta_b}_{ab}=\begin{pmatrix}
0&\langle n^T_a\rangle+\frac{1}{2}&0&0\\
\langle n^T_a\rangle+\frac{1}{2}&0&0&0\\
0&0&0&\langle n^T_b\rangle+\frac{1}{2}\\
0&0&\langle n^T_b\rangle+\frac{1}{2}&0\\
\end{pmatrix},
\end{equation}
from the properties of symplectic transformations we have
\begin{equation}
\bm{\sigma}^{\,\beta_a\,\beta_b}_{dc}= \bm{M}^{-1} \cdot \bm{\sigma}^{\,\beta_a\,\beta_b}_{ab} \cdot \left(\bm{M}^{-1}\right)^{T}.
\end{equation}
Finally we have
\begin{equation}
\langle \hat{d}^{\dagger}\hat{d}\rangle_{\beta_a \,\beta_b}=(\bm{\sigma}^{\,\beta_a\,\beta_b}_{dc})_{21}, \, \hspace{1cm} \langle \hat{c}^{\dagger}\hat{c}\rangle_{\beta_a \,\beta_b}=(\bm{\sigma}^{\,\beta_a\,\beta_b}_{dc})_{43}.
\end{equation}

\section{Entanglement}

We estimate the entanglement between the two modes via the logarithmic negativity of a two-mode Gaussian state \cite{Ferraro}.
At this aim we recall the position and momentum quadratures of the fluctuation operators of the two modes
\begin{equation}
\label{anncreaXY}
\begin{aligned}
&\hat{P}_x=\frac{1}{\sqrt{2}}\left(\delta\hat{a}^{\dagger}+\delta\hat{a}\right), \hspace{0.5cm} \hat{P}_y=\frac{i}{\sqrt{2}}\left(\delta\hat{a}^{\dagger}-\delta\hat{a}\right),\\
&\hat{A}_x=\frac{1}{\sqrt{2}}\left(\delta\hat{b}^{\dagger}+\delta\hat{b}\right), \hspace{0.5cm} \hat{A}_y=\frac{i}{\sqrt{2}}\left(\delta\hat{b}^{\dagger}-\delta\hat{b}\right),
\end{aligned}
\end{equation}
where $\hat{P}_i$ refers to the photons, $\hat{A}_i$ to the atoms $(i=x,y)$. In the case in which the first moments are null, as is it our case, the covariance matrix for the quadratures is defined as
\begin{equation}
\bm{S}_{ij}=\frac{1}{2}\langle \bm{\hat{u}}_i \bm{\hat{u}}_j + \bm{\hat{u}}_j \bm{\hat{u}}_i\rangle,
\end{equation}
with $\bm{\hat{u}}$ the vector $\bm{\hat{u}}=\left(\hat{P}_x, \hat{P}_y, \hat{A}_x, \hat{A}_y\right)^{\bm{T}}$.
It is useful to write the matrix explicitly as
\begin{equation}
\bm{{\cal S}}=\begin{pmatrix}
\bm{P}&\bm{X}&\\
\bm{X^T}&\bm{A}&\\
\end{pmatrix},
\end{equation}
where $\bm{X}$ refers to the correlations between the two modes. 
If we now introduce the quantity
\begin{equation}
\bm{\Sigma}(\bm{S})=\det \bm{P}+\det \bm{A}-2\det \bm{X},
\end{equation}
we can define
\begin{equation}
\nu_-=\frac{1}{\sqrt{2}}\,\sqrt{\bm{\Sigma}(\bm{{\cal S}})-\sqrt{\bm{\Sigma}(\bm{{\cal S}})^2-4\det \bm{{\cal S}}}}.
\end{equation}
The logarithmic negativity is then obtained as
\begin{equation}
E_N=\max\left(0,-\log2\nu_-\right),
\end{equation}
which is a measure of the quantum entanglement
, for a gaussian state defined by matrix $\bm{{\cal S}}$
, in the partition of modes $a$ and $b$. Analogously we can evaluate the entanglement in the partition of the polariton modes $d$ and $c$, via appropriate replacements of the relative operators.
\begin{figure}[t!]
\includegraphics[width=8cm]{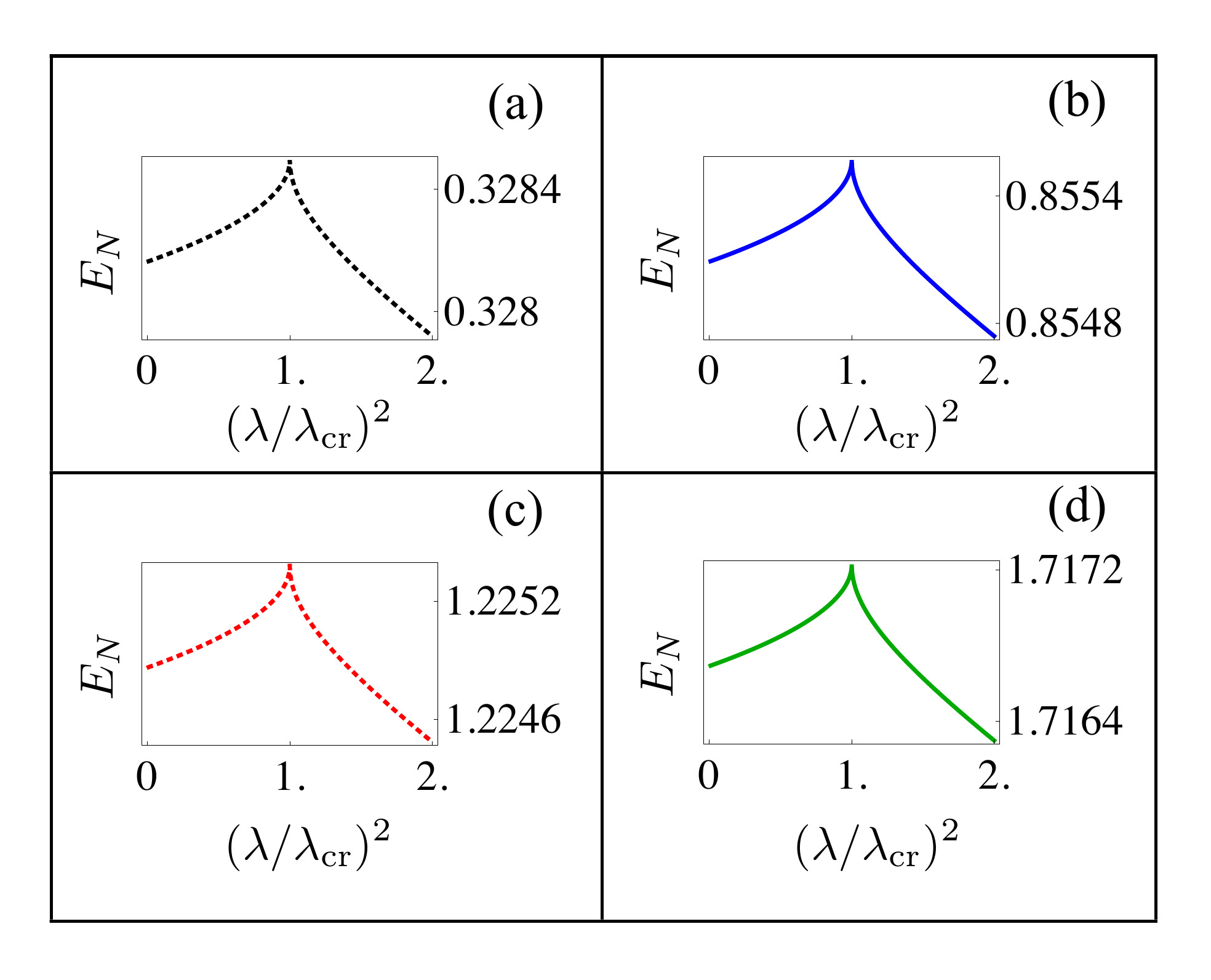}
\caption{Entanglement in the polariton partition $d$-$c$ for four locally passive entangled states, with different values of $\beta=1/K_B T$. Dotted black: $T=10^{-4}$,  Solid blue: $T=2\cdot 10^{-4}$, Dotted Red: $T=3\cdot 10^{-4}$, Solid Green: $T=4 \cdot10^{-4}$.}
\label{entanglement}
\end{figure}
In Fig.~\ref{entanglement} we show the entanglement for four locally passive entangled stats defined in the main text, for different values of the parameter $\beta$. These plots show that for the particular definition of this state, for a fixed value of $\beta$, the entanglement does not vary significantely with the coupling $\lambda$. In fact for our regime of parameter $\epsilon_- \gg \epsilon_+$ , and so the state is dependent almost only on $\epsilon_-$, which is almost independent on the coupling $\lambda$. This is particularly useful as it allows us to use the entanglement as a parameter, that increases as we go from panel (a) to (d), to analyse its role in the extraction of work.


\begin{thebibliography}{99}

\bibitem{nori} H.T. Quan et al. , \emph{Phys. Rev. E}, \textbf{76} 031105 (2007).

\bibitem{mahler} J. Birjukov, T. Jahnke, and G. Mahler, \emph{Eur. Phys. J. B} \textbf{64}, 105-118 (2008).

\bibitem{Friedenberger} A. Friedenberger, and E. Lutz, \emph{arXiv:1508.04128} (2015).

\bibitem{Uzdin} R. Uzdin, A. Levy, and R. Kosloff, \emph{Phys. Rev. X} \textbf{5}, 031044 (2015).

\bibitem{campisipekola} M. Campisi, J. Pekola, and R. Fazio, \emph{New J. Phys.} \textbf{17}, 035012 (2015).

\bibitem{campisisolo} M. Camipsi, \emph{Journal of Physics A: Mathematical and Theoretical} \textbf{47}, 245001 (2014).

\bibitem{kosloff} R. Kosloff, and A. Levy, \emph{Annual Review of Physical Chemistry}, \textbf{65} 365-393 (2014).

\bibitem{Kurizki} D. Gelbwaser-Klimovsky, R. Alicki, and G. Kurizki, \emph{EPL} \textbf{103}, 60005 (2013).

\bibitem{quan} H. T. Quan \emph{Phys. Rev. E} \textbf{79}, 041129 (2009).

\bibitem{AllahverdyanMahler} A. E. Allahverdyan, R. S. Johal, and G. Mahler, \emph{Phys. Rev. E} \textbf{77}, 041118 (2008).

\bibitem{Bruno} B. Leggio, B. Bellomo, and M. Antezza, \emph{Phys. Rev. A} \textbf{91}, 012117 (2015).

\bibitem{Bruno2} B. Leggio, and M. Antezza, \emph{arXiv:1601.08137} (2016).

\bibitem{Bruno3} P. Doyeux, B. Leggio, R. Messina, and M. Antezza, \emph{arXiv:1602.00031} (2016).

\bibitem{Felix} F. C. Binder, S. Vinjanampathy, K. Modi, and J. Goold, \emph{New J. Phys.} \textbf{17}, 075015 (2015).

\bibitem{Abah} O. Abah, J. Rossnagel, G. Jacob, S. Deffner, F. Schmidt-Kaler, K. Singer, and E. Lutz, \emph{Phys. Rev. Lett.} \textbf{109}, 203006 (2012).

\bibitem{Abah2} J. Ro\ss nagel, O. Abah, F. Schmidt-Kaler, K. Singer, and E. Lutz, \emph{Phys. Rev. Lett.} \textbf{112}, 030602 (2014).

\bibitem{single} J. Ro\ss nagel, S. T. Dawkins, K. N. Tolazzi, O. Abah, E. Lutz, F. Schmidt-Kaler, and K. Singer, \emph{arXiv:1510.03681}.

\bibitem{brazil} T. Batalh\~{a}o, A. M. Souza, L. Mazzola, R. Auccaise, I. S. Oliveira, J. Goold, G. De Chiara, M. Paternostro, and R. M. Serra, \emph{Phys. Rev. Lett.}  \textbf{113}, 140601 (2014).

\bibitem{An} S. An, J.-N. Zhang, M. Um, D. Lv, Y. Lu, J. Zhang, Z.-Q. Yin, H. T. Quan, and K. Kim, \emph{Nat. Phys.} \textbf{11}, 193-199 (2015).

\bibitem{Papero} M. Brunelli, L. Fusco, R. Landig, W. Wieczorek, J. Hoelscher-Obermaier, G. Landi, F. L. Semiao, A. Ferraro, N. Kiesel, T. Donner, G. De Chiara, and M. Paternostro, \emph{arXiv:1602.06958} (2016).

\bibitem{mauroextraction} M. A. Ciampini, L. Mancino, A. Orieux, C. Vigliar, P. Mataloni, M. Paternostro, and M. Barbieri, \emph{arXiv:1601.06796} (2016).

\bibitem{Marti} M. Perarnau-Llobet, K. V. Hovhannisyan, M. Huber, P. Skrzypczyk, N. Brunner, and A. Ac\'{i}n, \emph{Phys. Rev. X} \textbf{5}, 041011 (2015).

\bibitem{CampisiFazio} M. Campisi, and R. Fazio, arXiv:1603.05024 (2016).

\bibitem{DelCampo} J. Jaramillo, M. Beau, and A. del Campo, \emph{arXiv:1510.04633} (2015).

\bibitem{fialko} O. Fialko, and D. Hallwood, \emph{Phys. Rev. Lett.} \textbf{108}, 085303 (2012).

\bibitem{Azimi} M. Azimi, L. Chotorlishvili, S. K. Mishra, T. Vekua, W. Hübner, and J. Berakdar, \emph{New J. Phys.} \textbf{16},  063018 (2014).

\bibitem{Poletti} Y. Zheng, and D. Poletti \emph{Phys. Rev. E} \textbf{92}, 012110 (2015).

\bibitem{DickeETH} K. Baumann, C. Guerlin, F. Brennecke, and T. Esslinger, \emph{Nature} \textbf{464}, 1301 (2010).

\bibitem{Hardal} Ali \"{U}. C. Hardal, and \"{O}zg\"{u}r E. M\"{u}stecaplioglu, \emph{Sci. Rep.} \textbf{5}, 12953 (2015).

\bibitem{AllahverdyanThom} A. E. Allahverdyan, R. Balian, and T. M. Nieuwenhuizen, \emph{Entropy} \textbf{6}, 30-37 (2004).

\bibitem{Allahverdyan2004} A. E. Allahverdyan, R. Balian, and T. M. Nieuwenhuizen, \emph{EPL} \textbf{67}, 565 (2004).

\bibitem{GrossHaroche} M. Gross, and S. Haroche, \emph{Phys. Rep.} \textbf{93}, 301-396 (1982).

\bibitem{Vidal} J. Vidal, and S. Dusuel, \emph{Europhys. Lett.} \textbf{74}, 817 (2006).

\bibitem{Carmichael} H.J. Carmichael, C.W. Gardiner, and D.F. Walls, \emph{Phys. Lett. A} \textbf{46}, p.47 (1973).

\bibitem{PRLethDicke} K. Baumann, R. Mottl, F. Brennecke, and T. Esslinger, \emph{Phys. Rev. Lett.} \textbf{107}, 140402 (2011).

\bibitem{Brennecke} F. Brennecke, R. Mottl, K. Baumann, R. Landig, T. Donner, and T. Esslinger, \emph{PNAS} \textbf{110}, (29) 11763 (2013)

\bibitem{Dicke} R. H. Dicke, \emph{Phys. Rev.} \textbf{93}, 99 (1954).

\bibitem{Primakoff} T. Holstein and H. Primakoff, \emph{Phys. Rev.} \textbf{58}, 1098 (1949).


\bibitem{emary} C. Emary, and T. Brandes, \emph{Phys. Rev. A} \textbf{67}, 066203 (2003).

\bibitem{silva} F. N.C. Paraan, and A. Silva, \emph{Phys. Rev. E} \textbf{80}, 061130 (2009).

\bibitem{LutzWork} P. Talkner, E. Lutz, and P. H\"{a}nggi, \emph{Phys. Rev. E} \textbf{75}, 050102 (2007).

\bibitem{Marti2} N. Friis, M. Huber, and M. Perarnau-Llobet, arXiv:1511.08654 (2015).

\bibitem{Steve} G. Giorgi, and S. Campbell,  \emph{J. Phys. B: At. Mol. Opt. Phys.} \textbf{48}, 035501 (2015).

\bibitem{PlenioNEG} M. B. Plenio \emph{Phys. Rev. Lett.} \textbf{95}, 090503 (2005).

\bibitem{Horodecki} R. Horodecki, P. Horodecki, M. Horodecki, and K. Horodecki \emph{Rev. Mod. Phys.} \textbf{81}, 865 (2009).


\bibitem{Ferraro} A. Ferraro, S. Olivares, and M. G. A. Paris, ISBN 88-7088-483-X, Bibliopolis, Napoli, (2005).



\end{thebibliography}

\begin{thebibliography}{99}
\bibitem{emary} C. Emary, and T. Brandes, \emph{Phys. Rev. A} \textbf{67}, 066203 (2003).
\bibitem{Ferraro} A. Ferraro, S. Olivares, M. G. A. Paris, ISBN 88-7088-483-X, Bibliopolis, Napoli, (2005).
\end{thebibliography}
\end{document}